# Tracking Protection in Firefox
# For Privacy and Performance

Georgios Kontaxis
Columbia University
kontaxis@cs.columbia.edu

Monica Chew
Mozilla Corporation
monica.m.chew@gmail.com

*Abstract*—We present Tracking Protection in the Mozilla Firefox web browser. Tracking Protection is a new privacy technology to mitigate invasive tracking of users' online activity by blocking requests to tracking domains. We evaluate our approach and demonstrate a 67.5% reduction in the number of HTTP cookies set during a crawl of the Alexa top 200 news sites. Since Firefox does not download and render content from tracking domains, Tracking Protection also enjoys performance benefits of a 44% median reduction in page load time and 39% reduction in data usage in the Alexa top 200 news sites.

I. INTRODUCTION

Online advertising has a symbiotic relationship with the Internet ecosystem. Advertisers pay content publishers, i.e., websites, to embed promotional material in the content they generate. Publishers in turn use that revenue to mitigate the need for users to directly purchase the content they consume. In 2013, revenue from US online advertising reached $43 billion, supporting the vast majority of publishers [1]. A prime reason for its accelerated growth is the rise of efficient markets for targeted promotions.

Unfortunately, the technology to personalize advertisements and other site content relies extensively on tracking online activities of users. On top of that, governments misuse these technologies to facilitate warrantless surveillance without the knowledge or consent of the users or the original tracking services themselves.

The rise of online advertising has led to a corresponding increase in ad-blocking and privacy-oriented software. In 2013 ad-blocking grew nearly 70%, with nearly 41% of 18-29 year olds reporting usage [2]. Even though typical users may not be aware of the broad reach of tracking technologies, they feel that the advertising industry is insufficiently regulated and lack confidence protecting their data online [3], [4].

In response to this clarion call from users, we designed and implemented Tracking Protection in Mozilla Firefox 35. We demonstrate a 67.5% reduction in the number of HTTP cookies set during a crawl of the Alexa top 200 news sites. The reduction in cookies corresponds to blocking 11 tracking elements on 50% of these sites. Since Firefox does not download and render content from tracking domains, Tracking Protection also enjoys performance benefits of a 44% median reduction in page load time and 39% reduction in data usage in the Alexa top 200 news sites.

II. DESIGN AND IMPLEMENTATION

Tracking is a mechanism to identify a person and their browsing behavior. Tracking Protection for Firefox addresses a particular type of tracking that occurs when content from a tracking domain appears across multiple websites. Initially, tracking domains identified users through third-party cookies. Early attempts to mitigate tracking focused on blocking cookies, but other technologies utilizing local storage, fingerprinting, and etags evolved to be as powerful as cookies [5]–[7].

To cover the wide array of tracking technologies in use, Tracking Protection for Firefox prevents any network communication between the browser and unsafe third-party origins. It filters outgoing HTTP requests and cancels ones to known tracking domains. One challenge of this approach is that blocking content can break a site's functionality or appearance. For example, elements on the page may depend on a script being loaded from a tracking domain. To minimize unanticipated side-effects of Tracking Protection, we rely on a curated blocklist. Previous approaches have used heuristics or a default-deny policy at the expense of usability [8].

We implement an API based on Google Safe Browsing, a mechanism for efficient URL-based blocklist updates and lookups [9]. We use a subset of approximately 1500 domains from Disconnect's privacy-oriented blocklist to identify these unsafe origins [10]. We update the blocklist every 45 minutes to minimize the effects of incorrect blocklist entries.

When the browser is ready to make a network request, Firefox suspends it until Tracking Protection determines whether the destination URL is a tracking domain. If so, Firefox cancels the request before any data is sent or received. Canceling HTTP requests leaves the respective content loaders for documents, images, and scripts in a failed state. The network layer notifies individual loaders of cancellations and emits a document-wide security event.

Firefox indicates when it blocks tracking elements by showing a shield icon in the location bar (Figure 1). Clicking the shield presents a doorhanger that allows the user to disable Tracking Protection for the site in order mitigate the effects of false positives or loss of functionality. Disabling Tracking Protection updates the shield icon with a red strike-through while the doorhanger allows the user to re-enable protection (Figure 2).

To enable Tracking Protection in Firefox 35 and later, visit `about:config` and set `privacy.trackingprotection.enabled` to `true`.

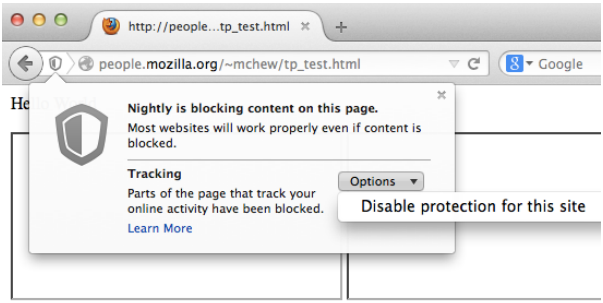

Fig. 1. When Tracking Protection blocks tracking elements, Firefox shows a shield icon in the location bar. Clicking the shield presents a dropdown panel that allows the user to disable Tracking Protection for that site.

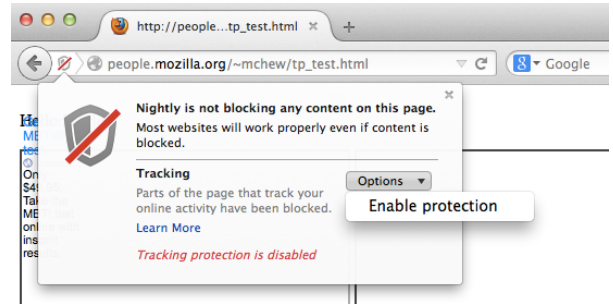

Fig. 2. If the user disables Tracking Protection for a site, Firefox shows a shield icon with a red strike-through to warn the user. Clicking the shield presents a dropdown panel that allows the user to re-enable protection.

TABLE I.  HTTP COOKIES SET BY THE ALEXA TOP 200 NEWS SITES. ENABLING TRACKING PROTECTION RESULTS IN 67.5% FEWER COOKIES.

| Configuration | # Cookies | Reduction |
|---|---|---|
| Default | 4006 | 0.0% |
| Tracking Protection | 1300 | 67.5% |
| Adblock Plus | 2398 | 40.1% |

## III. EVALUATION

We conduct simulated tests on Tracking Protection as well as measuring its effects in the field. We evaluate privacy benefits by measuring the reduction in HTTP cookies set and the number of blocked tracking elements per site, and the performance benefits by measuring reduction in page load time and data usage. We instrumented Firefox Nightly (the earliest pre-release version of Firefox) using Mozmill[1] to visit the Alexa 200 most popular news sites as our data set. We focus on news sites because they typify content providers who trade free content in exchange for advertising revenue. Each site was visited 10 times with and without Tracking Protection and the median values of identified trackers, load time and bytes downloaded were computed to represent it. Content caching and prefetching were disabled.

### A. Simulated testing

Tracking Protection blocks at least one unsafe element on 99% of the sites tested. In addition, Tracking Protection blocks 11 tracking elements in 50% of the sites and, in an extreme case, 150 tracking elements. Since tracking elements usually end up requesting more resources from tracking domains, these figures present only a lower bound of the number of elements that would load without Tracking Protection. Figure 3 shows the distribution of tracking elements per site. Table II lists the types of elements blocked by Tracking Protection.

To evaluate the effects of tracking on client-side state, we studied the browser's cookie store after crawling the Alexa top 200 news sites on three different configurations of Firefox Nightly: default (Tracking Protection disabled), Tracking Protection enabled, and Adblock Plus installed in a default lists. Table I shows Tracking Protection is more effective in reducing tracking cookies than Adblock Plus, resulting in a 67.5% reduction in cookies when compared to a default profile.

Figures 4 and 5 present the cumulative distribution function (CDF) of the reduction in median load time for the home page of each site and reduction in data usage, respectively. Load time is the interval between when the browser starts loading a top-level URL and when the page emits the `load` event. Data usage is the total number of bytes of all the resources under the top-level page. For each site, we measure load time and data usage with and without Tracking Protection enabled and calculate the reduction. The median reduction in page load time is 44%, and the median reduction in data usage is 39%. Studies show that users prefer faster load times [11], and conserving data can benefit bandwidth-constrained environments such as mobile. Note that the overhead eliminated by Tracking Protection comes from reducing third-party rather than first-party content on each page.

[1] https://developer.mozila.org/en-US/docs/Mozilla/Projects/Mozmill

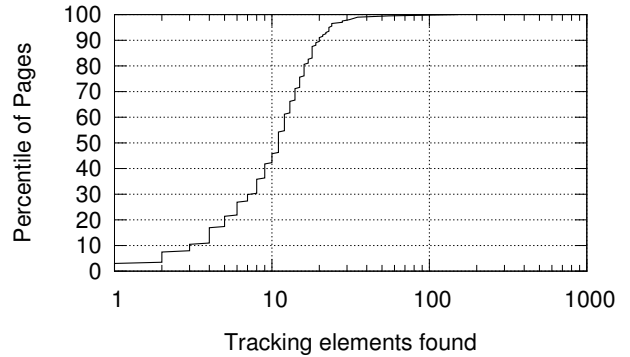

Fig. 3. CDF of the number of trackers on the Alexa top 200 news sites. 50% of them have at least 11 trackers present.

TABLE II.  DOM TYPES ASSOCIATED WITH TRACKING DOMAINS.

| Tracking Element | Count | Percent |
|---|---|---|
| SCRIPT | 1,890 | 79.2% |
| IMG | 334 | 14.0% |
| IFRAME | 149 | 6.3% |
| OBJECT | 13 | 0.5% |

As an example, www.weather.com loads in 3.5 seconds with Tracking Protection versus 6.3 seconds without and results in data usage of 2.8 MB (98 HTTP requests) versus 4.3 MB (219 HTTP requests), respectively. Even though Tracking Protection prevents initial requests for only 4 HTML `<script>` elements, without Tracking Protection, an additional 45 domains are contacted. Of the additional resources downloaded without Tracking Protection enabled, 57% are JavaScript (as identified by the `content-type` HTTP header) and, 27% are images. The largest elements appear to be JavaScript libraries with advertisement-related names, each on the order of 10 or 100 KB. Even though client-side caching can alleviate data usage, we observe high-entropy GET parameters that will cause the browser to fetch them each time. Privacy-conscious users clearing their cache often will also have to download them again.

TABLE III. TOP LEVEL PAGE LOADS FOR TRACKING PROTECTION.

| Page Loads | Count | Percent |
|---|---|---|
| Tracking Protection active (Figure 1) | 236,325 | 13.9% |
| Tracking Protection disabled (Figure 2) | 2,684 | 0.17% |
| No tracking elements present | 1,456,368 | 86% |
| Total | 1,695,557 | 100% |

*B. User testing*

From 2014 December 25 to 2015 January 7 inclusive, we used the Firefox Telemetry[2] framework to measure top-level page loads from Firefox Nightly users[3] who enabled Tracking Protection. Firefox Nightly enables Telemetry by default. Because of limitations in data collection, we do not know the exact number of users with Tracking Protection enabled. However, approximately 140,000 users run Firefox Nightly at least once a week, and during the test period approximately 0.5% of browser sessions enabled Tracking Protection. Table III presents a breakdown on page loads for sessions with Tracking Protection enabled. By policy Telemetry does not collect URL data, so multiple visits to the same website count as separate page loads. 14.1% of page loads contain at least one tracking element. Of these, users indicate breakage by disabling Tracking Protection only 1.2% (0.17%/13.9%) of the time. We can attribute the difference between the volume of page loads and number of sites containing at least one tracking element to the fact that popular sites account for a disproportionate number of page views.

## IV. RELATED WORK

Adblock Plus, Disconnect, and Ghostery are browser extensions utilizing curated, URL-based blocklists to prevent network requests for privacy-invasive or otherwise undesirable web content [10], [12], [13]. Adblock is the most popular add-on for Firefox and Chrome [14], but it incurs a prohibitively-high resource overhead [15]. What is more, users need to manually install these extensions as opposed to having that functionality built in the browser. Nevertheless, they are orthogonal to Tracking Protection in Firefox and can complement the set of tracking domains being blocked.

Privacy Badger is a browser extension that uses heuristics instead of a URL blocklist to identify and block traffic to tracking domains. Although preliminary research has been successful in algorithmically identifying tracking domains, operational experience with Privacy Badger suggests that relying on heuristics alone leads to an intolerable amount of site breakage for a typical user [8], [16], [17].

Microsoft Internet Explorer (MSIE) has had a feature similar to Firefox Tracking Protection, called Tracking Protection Lists (TPL), since 2011 [18]. Unlike Firefox Tracking Protection, Microsoft's TPL format supports allowing domains in addition to blocking them, with the allowlist taking precedence. Combining multiple lists can therefore lead to reduced coverage. MSIE also allows the user to select multiple lists, none of which are officially supported by Microsoft, leading to a confusing user experience [19].

## V. FUTURE WORK

Content trackers include user-visible components that may break functionality of the site if they are blocked. Examples of content trackers include social widgets such as the Facebook Like button, video hosting sites such as Vimeo, and commenting platforms such as Disqus. Research [16], [20] in this area prevents such content from loading until the user interacts with it, in which case the tracker is activated, or attempts to recreate the functionality in a privacy-preserving manner. For users who value content trackers, we must make the privacy-utility tradeoff clear and possibly offer the option to strip identifying information (cookies, etags) from outgoing network requests.

Another open challenge is applying Tracking Protection only to third-party content. We can avoiding cross-site tracking by blocking content from high-volume sites such as `facebook.com` without breaking them when visited directly. Heuristics such as the Public Suffix List[4] can help better determine the set of domains that are considered first-party. However, modern practices such as use of content distribution networks (e.g., Facebook's CDN is `fbcdn.net`) make relationships between sites difficult to discover.

Tracking Protection affects page rendering if blank areas appear in place of blocked tracking elements such as images or iframes. Some add-ons, such as Adblock Plus, attempt to resolve this problem by hiding these elements or removing them from the Document Object Model tree (DOM). However, these approaches may result in orphaned elements and other unattractive side-effects, including large performance penalties for re-drawing the DOM.

## VI. CONCLUSION

The Internet's principal revenue model leads to misaligned incentives between users, advertisers, and content providers, essentially creating a race to the bottom. Industry wields the most political and economic power, so it is up to users and user agents to advocate for the interests of people.

Do Not Track is a standards effort to communicate a user's preference regarding tracking to sites that they visit [21]. Despite having been in progress since 2009, DNT has not gained traction in industry, due to lobbying efforts from the ad industry. There are currently no legal or technological requirements tied to DNT. We cannot rely on the industry to self-regulate through standards efforts. A simpler approach may be to relinquish a standard definition of "tracking" and simply allow users to avoid visiting sites they don't find useful, regardless of that site's DNT policy.

Privacy-preserving advertising models such as Adnostic rely on behavioral targeting in the browser. We encourage this line of research, though it may not provide solace to users

---
[2] https://wiki.mozilla.org/Telemetry
[3] Nightly is the earliest pre-release channel and attracts the most technically proficient users.

[4] https://publicsuffix.org

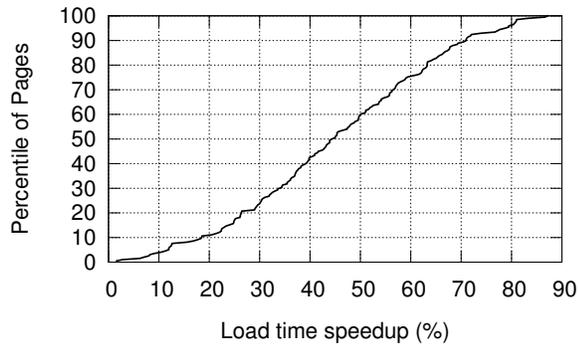

Fig. 4. CDF of the reduction in load time of the Alexa top 200 news sites. The median reduction in page load time is 44% with Tracking Protection.

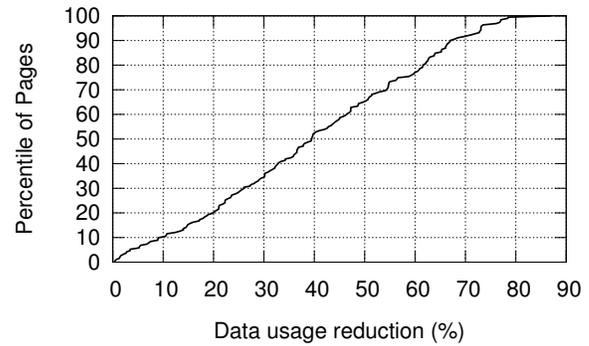

Fig. 5. CDF of the reduction in downloaded data of the Alexa top 200 news sites. The median reduction in data usage is 39% with Tracking Protection.

who find retargeting creepy, or the 30% of users who find ads annoying [2], [22].

At the same time, content providers who rely on advertising cannot stay in business without alternative revenue models. Google Contributor is a promising alternative revenue model for publishers. It allows people to pay a monthly fee between $1-3 for an ad-free experience, with part of their monthly contributions going to each site they visit [23].

Finally, browser makers bear tremendous responsibility in mediating conflicts between privacy interests of users and the advertising and publishing industries. Tracking Protection for Firefox is off by default and hidden in advanced settings. We call upon Mozilla, Microsoft, and other browser makers to make tracking protection universally available and easy to use. Only then will the balance of power shift towards interests of the people instead of industry.

A project page summarizing Tracking Protection in Firefox is available at https://kontaxis.github.io/trackingprotectionfirefox/.

ACKNOWLEDGMENTS

The authors would like to thank the team at Disconnect for their invaluable advice and making their list available to the public, and to the anonymous reviewers for their helpful comments. This work was supported in part by the National Science Foundation through Grant CNS-13-18415. Any opinions, findings, conclusions, or recommendations expressed herein are those of the authors, and do not necessarily reflect those of the US Government or the NSF. Part of the work of Georgios Kontaxis was done while at Mozilla.

REFERENCES

[1] Internet Advertising Bureau, "2013 Internet ad revenues soar to $42.8 billion," http://www.iab.net/about_the_iab/recent_press_releases/press_release_archive/press_release/pr-041014.

[2] PageFair and Adobe, "Adblocking goes mainstream," http://downloads.pagefair.com/reports/adblocking_goes_mainstream_2014_report.pdf.

[3] B. Ur, P. G. Leon, L. F. Cranor, R. Shay, and Y. Wang, "Smart, useful, scary, creepy: Perceptions of online behavioral advertising," in *Proceedings of the Eighth Symposium on Usable Privacy and Security*. ACM, 2012.

[4] Pew Research Center, "Public perceptions of privacy and security in the post-Snowden era," http://www.pewinternet.org/2014/11/12/public-privacy-perceptions/.

[5] G. Acar, C. Eubank, S. Englehardt, M. Juarez, A. Narayanan, and C. Diaz, "The web never forgets: Persistent tracking mechanisms in the wild," in *Proceedings of the 2014 ACM SIGSAC Conference on Computer and Communications Security*. ACM, 2014.

[6] K. Mowery and H. Shacham, "Pixel perfect: Fingerprinting canvas in HTML5," in *Proceedings of W2SP 2012*. IEEE Computer Society, 2012.

[7] P. Eckersley, "How unique is your web browser?" in *Proceedings of the 10th International Conference on Privacy Enhancing Technologies*. Springer-Verlag, 2010.

[8] Electronic Frontier Foundation, "Privacy Badger," https://www.eff.org/privacybadger.

[9] Google, "Safe Browsing API," https://developers.google.com/safe-browsing/.

[10] Disconnect, "Disconnect," https://disconnect.me.

[11] S. Lohr, "NYTimes - For impatient web users, an eye blink is just too long to wait," http://www.nytimes.com/2012/03/01/technology/impatient-web-users-flee-slow-loading-sites.html.

[12] W. Palant, "Ad Block Plus," https://adblockplus.org/.

[13] Ghostery, "Ghostery," https://www.ghostery.com.

[14] Mozilla, "Addons for Firefox," https://addons.mozilla.org/en-US/firefox/extensions/?sort=users.

[15] N. Nethercote, "AdBlock Plus's effect on Firefox's memory usage," https://blog.mozilla.org/nnethercote/2014/05/14/adblock-pluss-effect-on-firefoxs-memory-usage/.

[16] F. Roesner, T. Kohno, and D. Wetherall, "Detecting and defending against third-party tracking on the web," in *Proceedings of the 9th USENIX Conference on Networked Systems Design and Implementation*. USENIX Association, 2012.

[17] J. Bau, J. Mayer, H. Paskov, and J. C. Mitchell, "A promising direction for web tracking countermeasures," in *Proceedings of W2SP 2013*. IEEE Computer Society, 2013.

[18] Microsoft, "Tracking Protection Lists," http://ie.microsoft.com/testdrive/Browser/TrackingProtectionLists/Default.html.

[19] L. Cranor, "A first look at internet explorer 9 privacy features," https://www.techpolicy.com/Blog/March-2011/A-first-look-at-Internet-Explorer-9-privacy-featur.aspx.

[20] G. Kontaxis, M. Polychronakis, A. D. Keromytis, and E. P. Markatos, "Privacy-preserving social plugins," in *Proceedings of the 21st USENIX Conference on Security Symposium*. USENIX Association, 2012.

[21] Roy T. Fielding and David Singer, "Tracking preference expression (DNT)," http://www.w3.org/TR/tracking-dnt/.

[22] V. Toubiana, A. Narayanan, D. Boneh, H. Nissenbaum, and S. Barocas, "Adnostic: Privacy preserving targeted advertising," in *17th Annual Network and Distributed System Security Security Symposium*. The Internet Society, 2010.

[23] Google, "Contributor by Google," https://www.google.com/contributor/welcome/.